\documentclass[onecolumn,preprintnumbers,prl]{revtex4}
\usepackage{graphicx, subfigure}
\usepackage{amssymb, amsmath,amssymb,amsfonts}
\usepackage{amsthm,mathrsfs,amsopn}
\usepackage{dcolumn}
\usepackage{bm}
\usepackage{color}
\usepackage[utf8]{inputenc}

\theoremstyle{plain} 

\newcommand{\be}{\begin{equation}}
\newcommand{\ee}{\end{equation}}
\newcommand{\BE}{\begin{eqnarray}}
\newcommand{\EE}{\end{eqnarray}}
\newcommand{\BM}{\begin{multline}}
\newcommand{\EM}{\end{multline}}

\begin{document}

\title{Hopping in the crowd to unveil network topology}

\author{Malbor Asllani$^1$, Timoteo Carletti$^1$, Francesca Di Patti$^2$, Duccio Fanelli$^2$ and Francesco Piazza$^3$}
\affiliation{$^1$naXys, Namur Institute for Complex Systems, University of Namur, Belgium \\
$^2$ Dipartimento di Fisica e Astronomia, Universit\`a degli Studi di Firenze and INFN, Sezione di Firenze, Italy\\
$^3$ Centre de Biophysique Mol\'eculaire, CNRS-UPR 4301, Universit\'e d'Orl\'eans, France
}

\begin{abstract}
We introduce a nonlinear operator to model  diffusion on a complex undirected network under crowded conditions. 
We show that the asymptotic distribution of diffusing agents is a nonlinear function of the nodes' degree and saturates 
to a constant value for sufficiently large connectivities, at variance with standard diffusion
in the absence of excluded-volume effects. Building on this observation, we define and solve an inverse problem, 
aimed at reconstructing the a priori unknown connectivity distribution. 
%
%
The method gathers all the necessary information by repeating a limited number of independent 
measurements of the asymptotic  density  at a single node that can be chosen randomly.
The technique is successfully tested against both synthetic and real data, and shown to 
estimate with great accuracy also the total number of nodes.  
\end{abstract}

\maketitle

Networks are everywhere. The brain, Internet and the cyberworld, foodwebs, social contacts 
and commuting fall within the vast realm of network science~\cite{newman,latora,albert}. Irrespective 
of the specific domain of application, individual entities (e.g. material units, bits of information)  
belonging to a certain population, may jump from one site (node) to its adjacent neighbors, 
following the intricate web of distinct links, which defines the architecture 
of a complex network~\cite{vespignani}. The ability of agents to explore the network to which they are bound is customarily 
modeled as a standard diffusive process~\cite{vespignani,sneppen}. Individuals are therefore independent from each other,
while mutual interference stemming from the competition for the available space  is 
deliberately omitted. However, in real-world applications the carrying capacity of each node is finite, 
an ineluctable constraint that should be accommodated for under crowded 
conditions~\cite{ligget,almaas,kwon,fanelli,GalantiCrowding,Fernando:2010fk,Landman:2011uq,Galanti:2014zr,Galanti2}. 
On the other hand, the structure of the network is often unknown. Several methods have been introduced 
in the literature aimed at reconstructing the topology of the network from a direct assessment 
of the transport dynamics~\cite{nature,pnas,livi}. This amounts to solving an inverse problem, 
from functions back to structure, a task which involves formidable challenges \cite{timme,livi,irene}. \\
\indent Working along these lines, the aim of this paper is twofold. On the one side, 
we will introduce a nonlinear operator to describe diffusion in a crowded network. 
The derivation originates from a microscopic stochastic framework and it is solidly grounded 
on first principles. As an important byproduct of the analysis, we will discuss the idea 
of {\em functional degree centrality}, as opposed to the usual structural notion that is
routinely invoked in network studies. Then, we will outline an innovative scheme 
which  exploits the dynamical entanglement among walkers
to reconstruct the unknown topology of the network. The method allows one 
to accurately determine the distribution of connectivities (degrees) and to 
quantify, with unprecedented efficiency, the size of the examined network, 
as we shall here demonstrate for a selection of paradigmatic examples.
At variance with other available approaches, in our method one 
gathers all necessary information from just one random node of the collection. \\
\indent Let us consider a generic undirected graph composed of $\Omega$ nodes and 
characterized by its adjacency matrix $\textbf{A}$
($A_{ij}=1$, if nodes $i$ and $j$ are connected, zero otherwise). 
The degree of each node, namely the number of connected neighbors, 
is $k_i=\sum_j A_{ij}$. Each node is endowed with a given {\em carrying capacity}, i.e. 
is assumed to be partitioned into a large number $N$ 
of compartments, which can be either occupied by an agent 
(a physical or abstract entity) or empty. The stochastic dynamics of a 
collection of particles randomly hopping on 
a network and competing for the  the available space within nodes 
is described by the following master equation
\begin{widetext}
\begin{equation}
\frac{d}{dt}P(\mathbf{n},t)=\sum_{i,j} A_{ij} ( T(n_i,n_j| n_i+1,n_j-1) P(n_i+1,n_j-1,t) 
                                              -   T(n_i-1,n_j+1| n_i,n_j) P(n_i,n_j,t) ) 
\label{eq:master}
\end{equation}
\end{widetext}
where $P(\mathbf{n},t)$ denotes the probability that the system be in the 
state $\mathbf{n}=(n_1,\dots , n_\Omega)$ at time $t$ \cite{fanelli}. 
The scalar quantity $n_i$ identifies the number of particles on node $i$.
In the  right-hand side of Eq.~\eqref{eq:master} we indicate explicitly 
only the species which are involved in the selected transition.
In this stochastic process, one of the walkers sitting on node $i$ is selected
with probability $n_i/N$ and then made to jump to one of the $k_i$ neighboring nodes
with  probability $1/k_i$. However, this can only occur if the target node is not fully occupied,
i.e. with probability $\left (N - n_j \right )/N$. 
Overall, the hopping probability $T( \cdot | \cdot )$ from  node $i$ to node $j$ 
reads
\begin{equation}
T(n_i-1,n_j+1| n_i,n_j)=\frac{1}{k_i} \frac{n_i}{N} \frac{ \left (N - n_j \right )}{N}.
\end{equation}
The above nonlinear transition probabilities accommodate for excluded-volume interactions \cite{ligget}, i.e.
the jump rate is modulated by the crowding at destination. 
Multiplying by $n_i$  both sides of Eq.~\eqref{eq:master}, and summing over all $n_i$~\cite{vankampen}, 
one obtains the rate equations for the evolution of the average density
$\langle n_i\rangle \equiv \sum_{\mathbf{n}} n_i P(\mathbf{n},t)$. 
In the thermodynamic limit $\rho_i=\lim_{N\rightarrow\infty}\langle n_i\rangle/N$ 
and a straightforward manipulation yields
\begin{equation}
\frac{\partial }{\partial t}\rho_i=\sum_{j=1}^\Omega \Delta_{ij}\left[\rho_j\left(1-\rho_i\right)-\frac{k_j}{k_i}\rho_i\left(1-\rho_j\right)\right] \label{eq:diffusionDelta}
\end{equation}
where use has been made of the fact that $A_{ij}=A_{ji}$ and  
$\langle n_in_j\rangle=\langle n_i \rangle \langle n_j \rangle$
(a condition that holds exact when $N\rightarrow\infty$ \cite{vankampen}). 

\begin{figure*}[t!]
\centering
\includegraphics[width=0.8\textwidth]{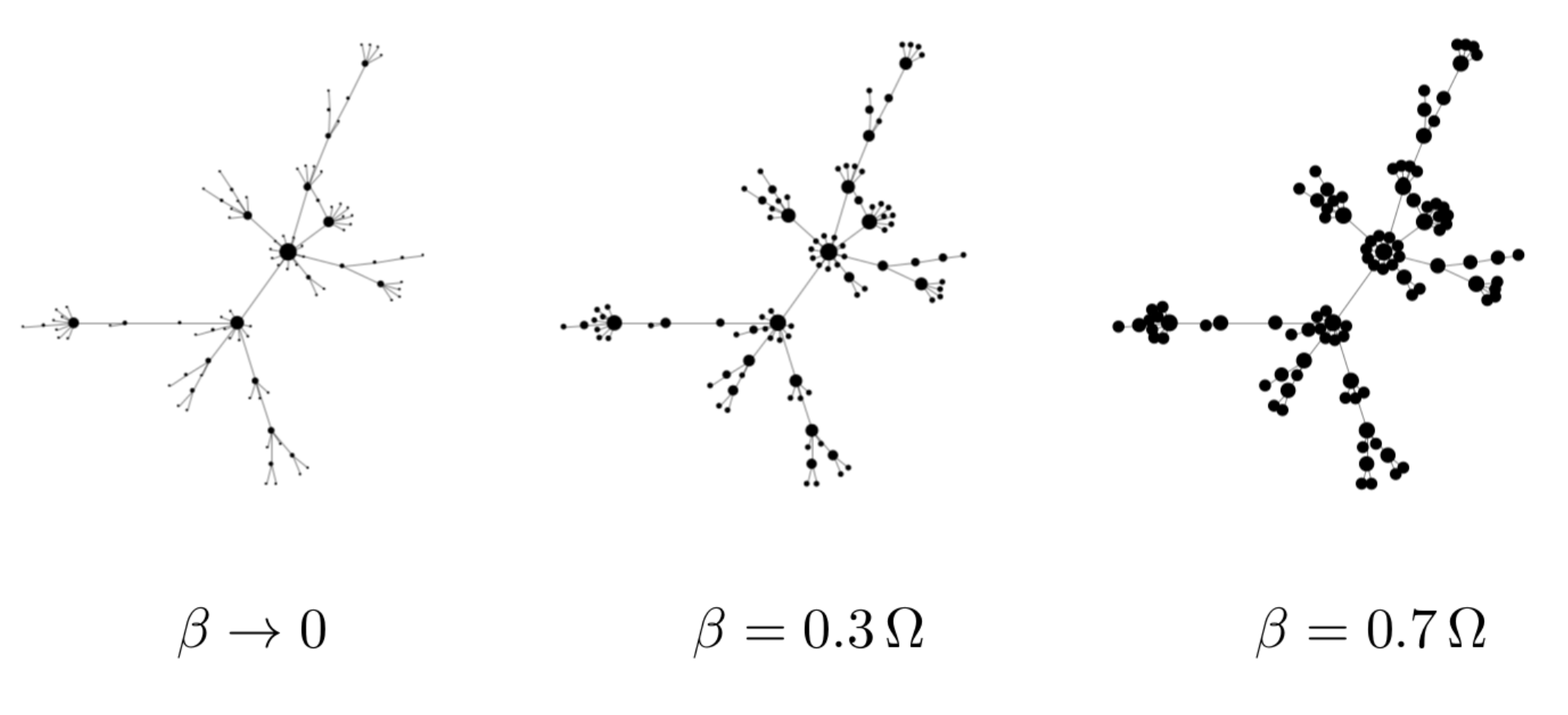}
\caption{Asymptotic distribution of  walkers diffusing on a scale-free network under different 
crowding conditions, as specified by $\beta$. Standard diffusion is recovered in the 
limit $\beta \rightarrow 0$ and displayed in the leftmost picture. 
Nodes are drawn with a size proportional to the corresponding asymptotic density, $\rho_i^{\infty}$.}
\label{fig:Fig1}
\end{figure*}

\begin{figure*}[ht!]
\begin{centering}
\begin{tabular}{ll}
\textbf{a} & \textbf{b}\\
\includegraphics[scale=0.4]{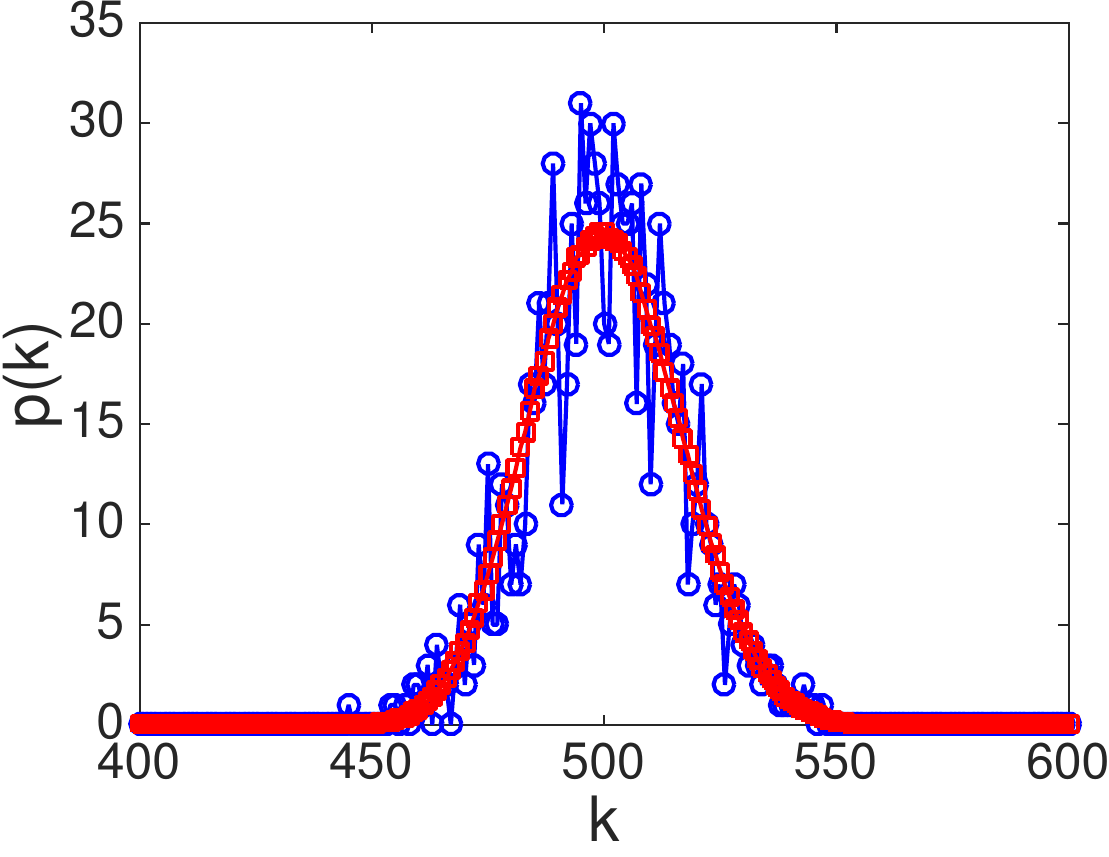} &
\includegraphics[scale=0.4]{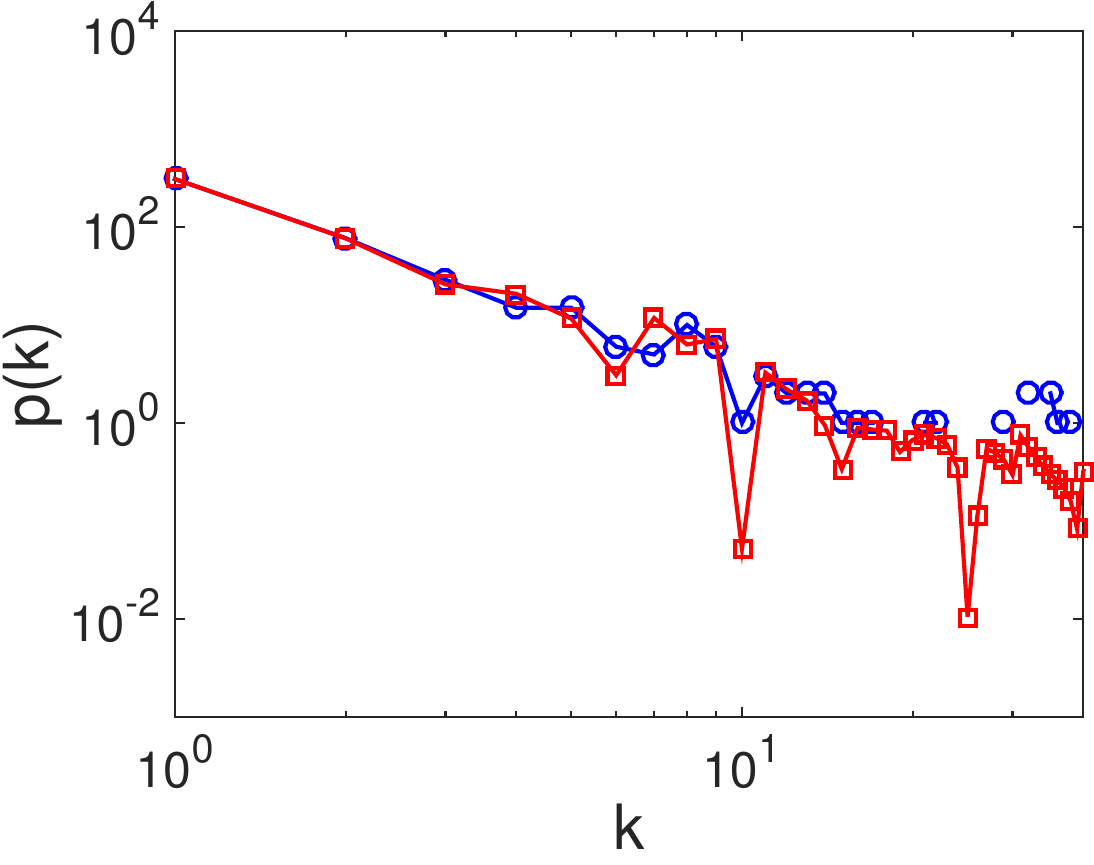} \\
\textbf{c} & \textbf{c} \\
\includegraphics[scale=0.4]{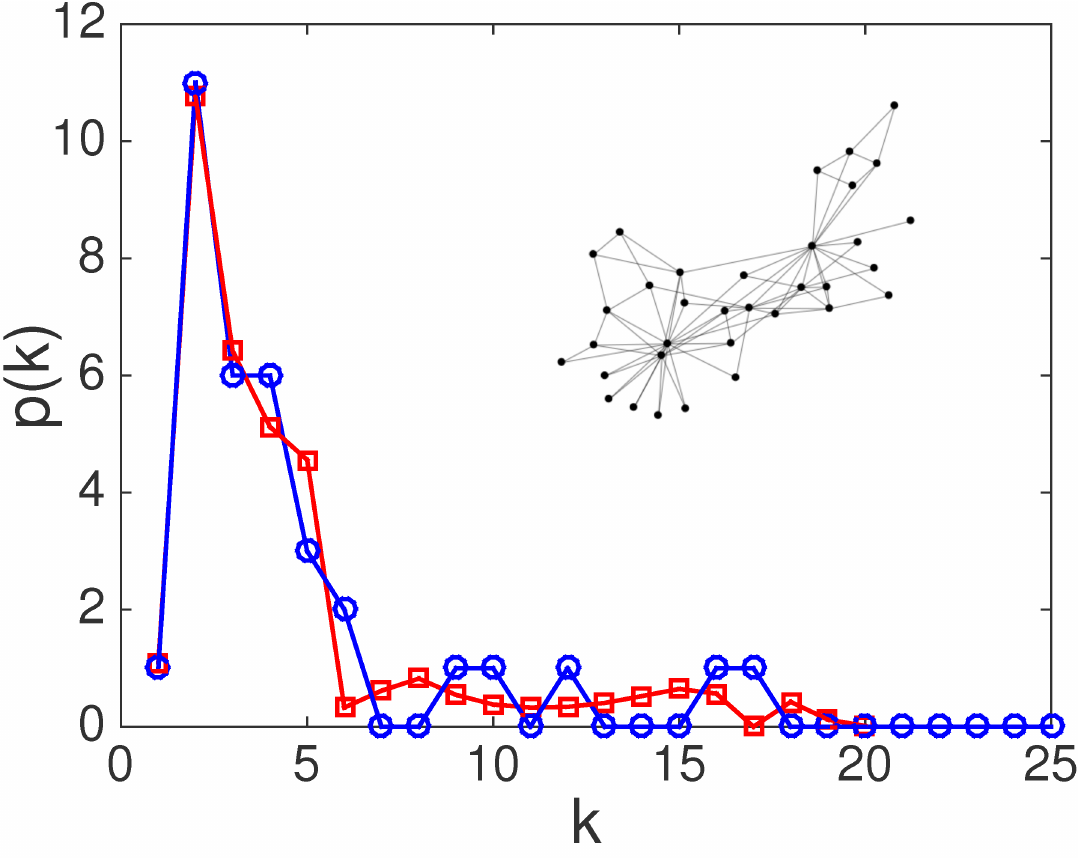} &
\includegraphics[scale=0.4]{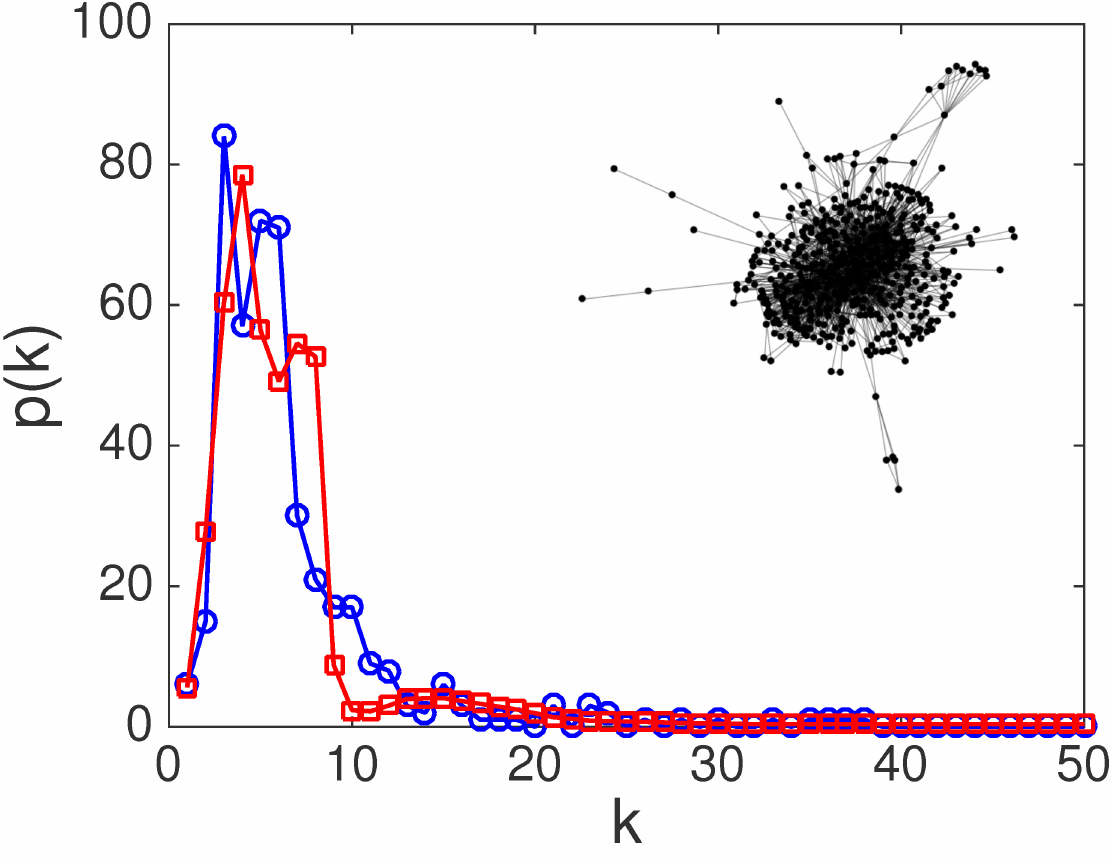} 
\end{tabular}
\end{centering}
\caption{\label{fig:Fig2} Examples of network reconstructions, $p(k)$ vs $k$. Panel (a): 
Erdos-Renyi network  with $\Omega=1000$. The probability $p$ for a link between 
any two pairs of nodes is set to $0.5$. Exact profiles (blue circles) and profiles reconstructed with $s=600$ (red squares).
The estimated number of nodes is $\Omega=\sum_k p(k) = 1000$. 
Panel (b): scale-free network with $\Omega=500$ and $\gamma=2$. The degree distribution $p(k)$ 
is plotted in log-log scale, focusing on the relevant region in $k$. 
Blue circles represent the exact distribution, while red squares identify the profile reconstructed 
with $s=150$ independent single-node measurements. The estimated number of nodes 
is $\Omega=\sum_k p(k) = 500.2$. Panel (c):  reconstructed (red squares) versus exact (blue circles) 
degree distribution for the undirected Karate club network (shown in the inset),
$\Omega=34$. Here, $s=20$. The estimate size of the network is $\Omega=\sum_k p(k) = 34.02$. 
Panel (d): the performance of the method is assessed by using the (symmetrized) C. elegans 
metabolic network (displayed in the inset) with $\Omega=453$.
Blue circles depict  the exact distribution, while red squares refer 
to the reconstruction performed with $s=200$ experiments, which gives $\Omega=\sum_k p(k) = 453.76$.
\label{fig:instabilityRegion}}
\end{figure*}

In the above equation $\Delta_{ij}=A_{ij}/k_j -\delta_{ij}$ denotes the elements 
of the Laplacian $\mathbf{\Delta}$, the transport operator  which describes 
diffusion of non-interacting agents on a heterogeneous network~\cite{sneppen}. 
Working under diluted conditions amounts to neglecting  nonlinear terms 
in Eq.~\eqref{eq:diffusionDelta}, which  therefore reduces
to the standard diffusion equation 
$\dfrac{\partial }{\partial t}\rho_i=\sum_{j=1}^\Omega \Delta_{ij}\ \rho_j$~\footnote{As a side remark 
we observe that Eq. (\ref{eq:diffusionDelta}) can be cast in the equivalent form
$\frac{\partial }{\partial t}\rho_i=\sum_{j=1}^\Omega \Delta_{ij}\rho_j - \rho_i \sum_{j=1}^\Omega \Delta_{ij}\rho_j + \frac{\rho_i}{k_i} \sum_{j=1}^\Omega \Delta_{ij}\rho_jk_j$. When ($k_i=k$ $\forall i$), as it happens on a regular 
lattice subject to periodic boundary conditions, the last two terms cancel out and one recovers the linear diffusion equation. Stated differently,  the signature of the microscopic exclusion rule disappear 
when heterogeneity gets lost. This is the generalization of an observation made by Huber in~\cite{Huber}, 
see also~\cite{GalantiCrowding}.}. %
The mass (total number of individuals) is an invariant of the dynamics, which means that the quantity 
$\beta=\sum_{i=1}^\Omega\rho_i(t) \in (0,\Omega]$ is  conserved.\\
\indent The equilibrium solution $\rho_i^{\infty}$ of Eq.~\eqref{eq:diffusionDelta} can be determined analytically  
following a simple reasoning. A stationary solution of Eq.~\eqref{eq:diffusionDelta} 
can be found by setting $\rho_j^{\infty}\left(1-\rho_i^{\infty}\right)-\rho_i^{\infty} 
k_j/k_i \left(1-\rho_j^{\infty}\right)=0$, $\forall \, i,j$. 
Disregarding the trivial solution $\rho_i^{\infty}=1$, 
one gets $\rho_j^{\infty}= a_i k_j/(1+a_ik_j)$, where
$a_i=\rho_i^{\infty} \left [ k_i\left(1-\rho_i^{\infty}\right) \right ] ^{-1}$. 
By definition, $\rho_j^{\infty}$ should not depend on $i$, which in turn implies $a_i=a$ $\forall \,i$.
This gives~\footnote{The same conclusion can be reached by plugging (\ref{eq:stat}) into the 
governing dynamical equation (\ref{eq:diffusionDelta})  and showing that it yields $\dot \rho_i =0 $ $\forall i$.}
\be
\rho_i^\infty=\frac{ak_i}{1+ak_i}
\label{eq:stat}
\ee
where the constant $a$ should be  determined from the normalization condition 
$\sum_{i=1}^\Omega (ak_i)/(1+ak_i) = \beta$. It can be shown that Eq.~\eqref{eq:stat} 
is the only  non-trivial equilibrium for system~\eqref{eq:diffusionDelta}.  
Under diluted conditions, Eq.~\eqref{eq:stat} returns the standard equilibrium 
solution for the diffusion operator: the asymptotic distribution of walkers at node $i$ is
proportional to its connectivity, $k_i$. 
Peaks in the steady-state concentration identify therefore structural hubs of the network, 
while peripheral nodes are associated with modest densities. 
In many practical applications, including page ranking schemes, the  distribution of 
hopping walkers is hence believed to return an immediate measure of the nodes' centrality. 
Further,  punctual information as delivered by a large, although finite, population of individual 
diffusing on a network, is usually processed by assuming a distribution of the incoming signals
which scales linearly with the connectivity of the nodes~\cite{newman}. 
Nonlinear interference among interacting walkers competing for space 
alters significantly the aforementioned simple scenario. For sufficiently large connectivities, 
the predicted distribution~\eqref{eq:stat} reaches a constant value. This effect is more pronounced 
the larger the value of $a$ or, equivalently, the larger $\beta$. 
In Figure \ref{fig:Fig1} we report the equilibrium density  obtained from 
Eq.~\eqref{eq:stat} on a scale-free network under 
different crowding conditions. When $\beta$ grows, hubs become progressively less distinct. 
{\it Structural} centrality, as revealed by the distribution of non interacting agents 
(and asymptotically approached in the limit $\beta \rightarrow 0$), differs 
significantly from {\it functional} centrality, which follows the equilibrium 
distribution when $\beta \ne 0$ and nodes are assigned a finite carrying capacity. 
Elaborating on this dichotomy is particularly relevant for scale-free networks.\\
%
%
\indent As anticipated, the nonlinear transport operator introduced above opens up the perspective 
of designing a novel scheme to access key global features of a network 
from direct measurements of the steady-state dynamics. We aim in particular at determining  
$p(k)$, the distribution of connectivities $k$, which conditions implicitly the equilibrium 
density of agents~\footnote{$p(k)$ is normalized so as to yield $\sum_k p(k) =\Omega$.}. 
Importantly, our method also allows us to estimate the total number of nodes  
$\Omega$ in the network. In the following, we will outline the mathematical 
steps of the procedure and then move forward to discussing a selected gallery of case studies.\\
\indent For any fixed $\beta$, we monitor the dynamics of the system at {\it just one node} 
of the collection, hereafter $i$. After a sufficiently long time, one can 
measure the local asymptotic concentration $\rho_i^{\infty}$. Assuming the local connectivity 
$k_i$ to be known, which is a reasonable working hypothesis given that we are sitting 
at node $i$, we can write
\begin{equation*}
a(\beta)=\frac{\rho_i^{\infty}}{1-\rho_i^{\infty}}\frac{1}{k_i}\, ,
\end{equation*}
where the dependence of $a$ on $\beta$ has been emphasized. 
We now rewrite the normalization condition 
so as to bring  $p(k)$ into the picture, that is, 
\begin{equation}
\beta=\sum_k p(k)\frac{a(\beta)k}{1+a(\beta)k}\, .
\label{mass_cons}
\end{equation}
Since the network is limited in size, the previous sum involves a finite number of 
terms. Performing $s$ independent measurements (or experiments), carried out 
for different values of the constant $\beta$, yields
\begin{equation}
\label{eq:betaFp}
\bm{\beta}=\mathbf{F} \mathbf{p}\, ,
\end{equation}
where $\bm{\beta}=(\beta_1,\dots,\beta_s)^T$, 
$\mathbf{p}=(p(1),\dots,p(s))^T$ and $F_{lr}$, 
the generic element of matrix $\mathbf{F} $, is given by
\begin{equation*}
F_{lr}=\frac{ra(\beta_l)}{1+ra(\beta_l)}\, .
\end{equation*}
Determining the components of the vector $\mathbf{p}$ amounts to inverting matrix $\mathbf{F}$. Notice that $s$, 
the number of independent measurements performed, should be at least equal to the maximum degree 
of the inspected network, which is not known a priori. As it is shown in the SI, one can progressively 
increase $s$ until the reconstructed $p(k)$ converges to a stable profile. 
Monitoring the first moments of the 
distribution against $s$ provides a robust stopping criterion (see SI).\\
\indent The matrix $\mathbf{F}$ can be ill-conditioned when the size of the 
inspected network is too large. In this case, dedicated regularization schemes can be employed  
to carry out the matrix inversion and recover the degree distribution~\cite{golub}. 
To further improve the accuracy of the reconstruction, 
one can also impose additional constraints on the inversion algorithm, such as requiring a positive-defined $p(k)$. 
A comment is in order before turning to test the adequacy of our methodology. 
Our protocol to reconstruct the unknown distribution of connectivities  works 
if excluded-volume effects are  accounted for. Under extremely diluted conditions,  
standard diffusion holds and $\rho_i^\infty = b k_i$, with $b$ constant. 
In this case,  Eq.~\eqref{mass_cons} yields the trivial condition 
$\langle k \rangle=\beta/b$, where $\langle k \rangle=\sum_j p(k_j) k_j$. 
Hence, performing many experiments for different values of $\beta$ 
simply returns independent estimates of the first moment of the distribution. 
It is the nonlinear nature of~\eqref{eq:stat}, the macroscopic blueprint of crowding, which 
enables the higher order moments of $p(k)$ to be self-consistently computed 
from independent experiments. In the remaining part of this paper, 
we will validate the procedure against a limited selection of case studies
that bear both theoretical and applied interest.\\
\indent As a first example, we consider an Erdos-Renyi random network~\cite{gilbert} with
$\Omega=1000$ nodes and a probability for two random nodes to be linked $p=0.5$. 
According to the procedure discussed above, we track the evolution of the 
walkers and measure the asymptotic density at a generic node $i$ whose  connectivity $k_i$ 
is assumed to be known. Here, $s=600 < \Omega$ independent experiments are performed and the 
information processed to infer the distribution of connectivity $p(k)$. 
In Figure~\ref{fig:Fig2}(a) we compare the reconstructed profile with 
the exact distribution. The agreement is good and illustrates well the efficacy of our procedure.  
Integrating the mean-field deterministic dynamics, or operating under the original stochastic 
framework~\cite{gillespie}, returns consistent and equally accurate results (see SI). 
From the reconstructed distribution  one can readily calculate $\Omega=\sum_k p(k)$ 
the total number of nodes of the network. In this case, 
with $s=600$ mesurements one obtains the true value $\Omega=1000.0$.\\
%
%
\indent The second example is a scale-free network generated according to the preferential 
attachment rule~\cite{barabasi}. In Figure~\ref{fig:Fig2}(b) we compare  
the exact distribution of connectivities (circles) 
with the approximated solution (squares) inferred through our algorithm. 
Also in this case the agreement is satisfying: the power-law  scaling is correctly 
reproduced and the predicted number of nodes in excellent agreement with the true value (see caption).\\
\indent Finally, we turn to study two other examples that build on real networks:  
the well known Karate Club network~\cite{karate,konect}  and the C. elegans
metabolic network (which  was artificially made undirected)~\cite{elegans,konect}. 
Diffusion  is assumed to occur subject to finite-volume constraints and the method exemplified above 
is applied in order  to recover the degree distribution from single-node measurements of 
the steady-state. The results reported in Figure~\ref{fig:Fig2}(c) and~\ref{fig:Fig2}(d), respectively, 
quantify the  predictive power of the proposed methodology.\\
\indent Summing up, we have  introduced a novel nonlinear operator to model 
the process of diffusion on a complex network under crowded conditions. 
Agents may randomly hop from one node to another, provided free space is available at the target 
destination. The asymptotic density distribution in the presence of crowding 
differs significantly from that predicted by standard diffusion, i.e.  under diluted conditions. 
In the latter case, the density scales linearly with the nodes' degree, 
while in crowded conditions the equilibrium concentration saturates for large enough values of the
connectivity. Based on this observation, we discussed the notion of functional hubs,  
as opposed to that based only on topology.
Further, we developed an efficient scheme that enables one to infer the unknown 
distribution of connectivities from single-node measurements of the asymptotic diffusion dynamics.
The method takes advantage of the interference that builds up among microscopic agents
as a consequence of the competition for the available space. 
Tests are performed using both synthetic and real networks, which illustrate convincingly the power of 
our method. \\
\indent In its present formulation our approach is suited for symmetric networks only.
Future extensions are planned to reconstruct the distribution of outgoing and incoming connectivities 
in asymmetric crowded networks, by examining the directed flux of particles from just one node. 
Finally, it would be interesting to gauge the impact of endowing nodes with a finite carrying capacity 
on other  measures of centrality introduced in the literature. 

\section*{Acknowledgments}
We thank Renaud Lambiotte for insightful comments.
The work of M.A. is supported by a FRS-FNRS Postdoctoral Fellowship. 
The work of T.C. presents research results of the Belgian Network DYSCO 
(Dynamical Systems, Control, and Optimization), funded by the Interuniversity 
Attraction Poles Programme, initiated by the Belgian State, Science Policy Oce. 
The scientific responsibility rests with its author(s).  
D.F. acknowledges financial support from H2020-MSCA-ITN-2015 project COSMOS  642563. 

\newpage
\appendix
\section*{Supplementary Information. Hopping in the crowd to unveil network topology }

\section{On the convergence of the reconstruction scheme.}

\begin{figure*}[ht!]
\begin{centering}
\begin{tabular}{llll}
\textbf{a} & \textbf{b} & \textbf{c} & \textbf{d} \\
\includegraphics[scale=0.2]{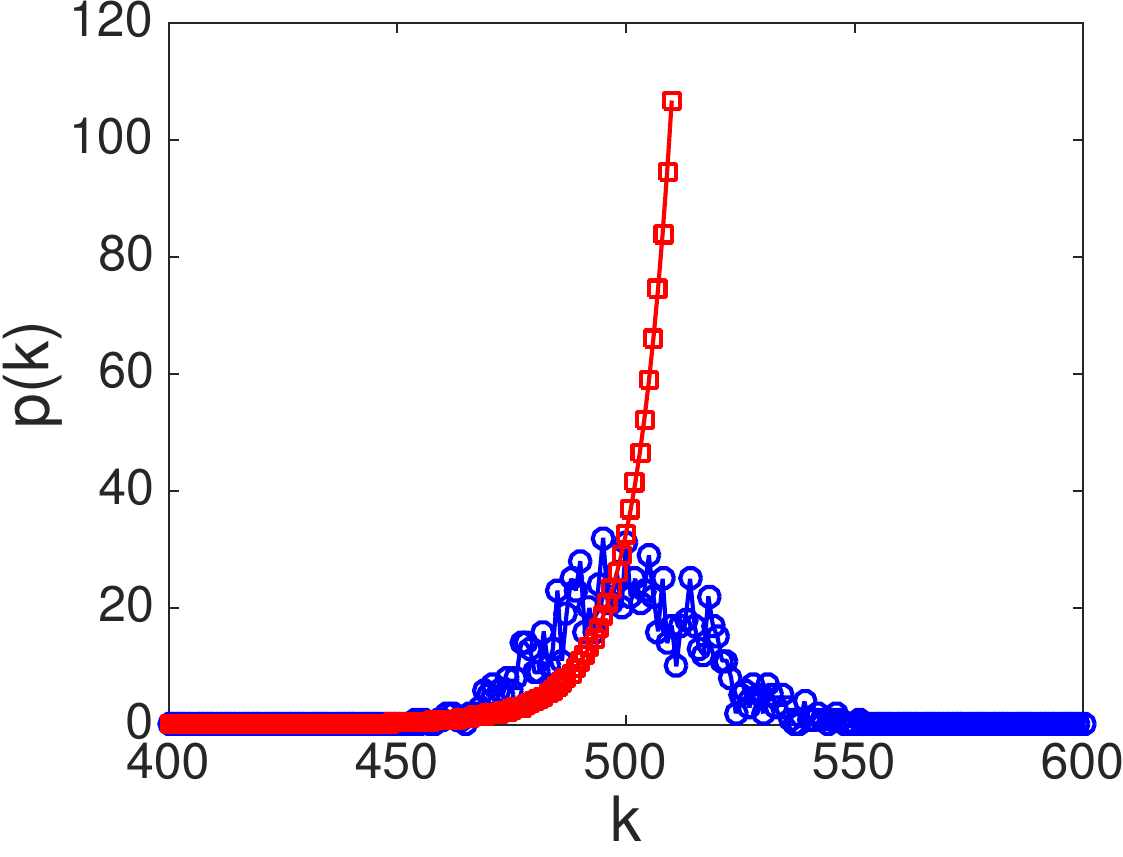} & \includegraphics[scale=0.2]{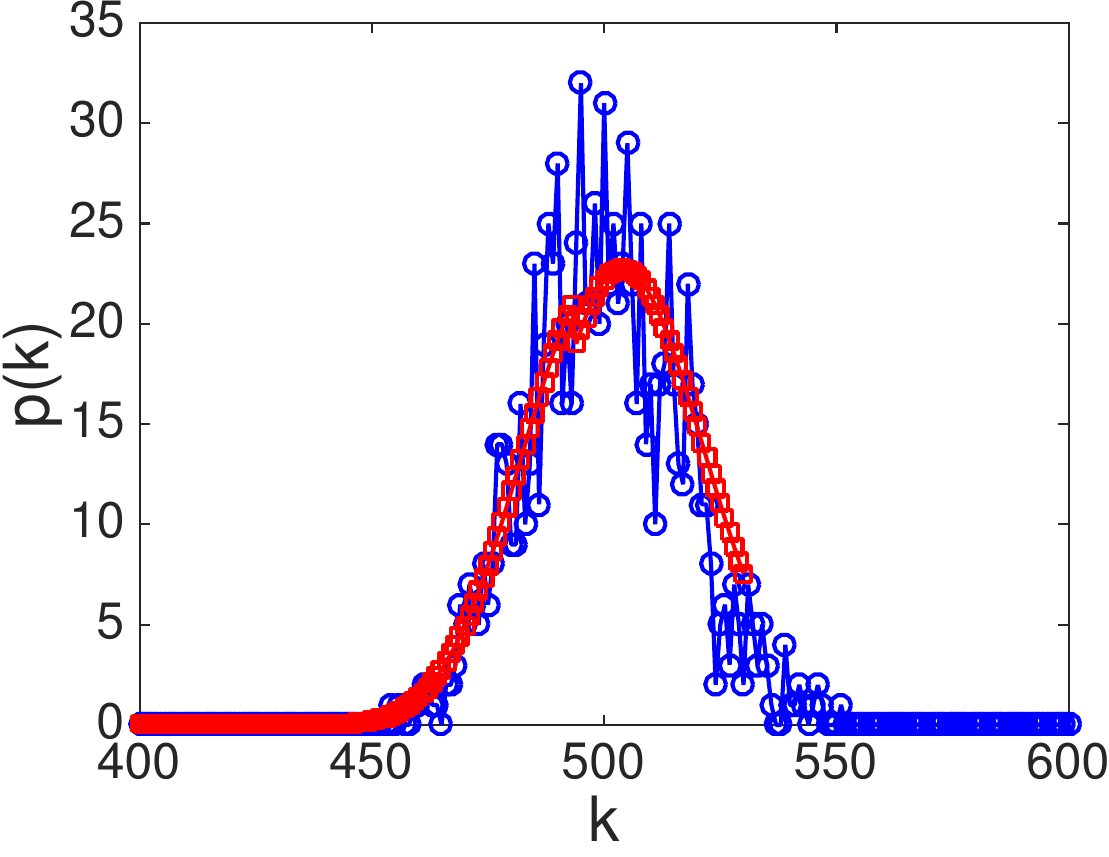} & \includegraphics[scale=0.2]{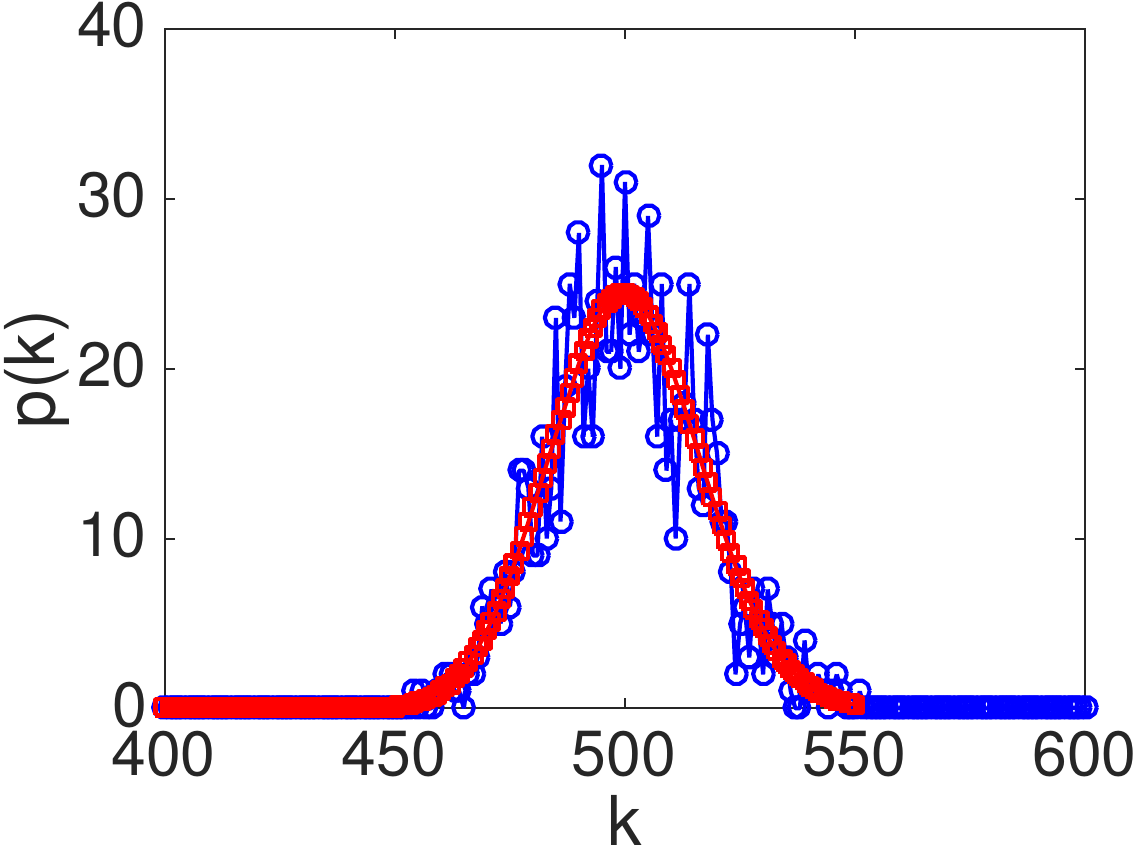} & \includegraphics[scale=0.2]{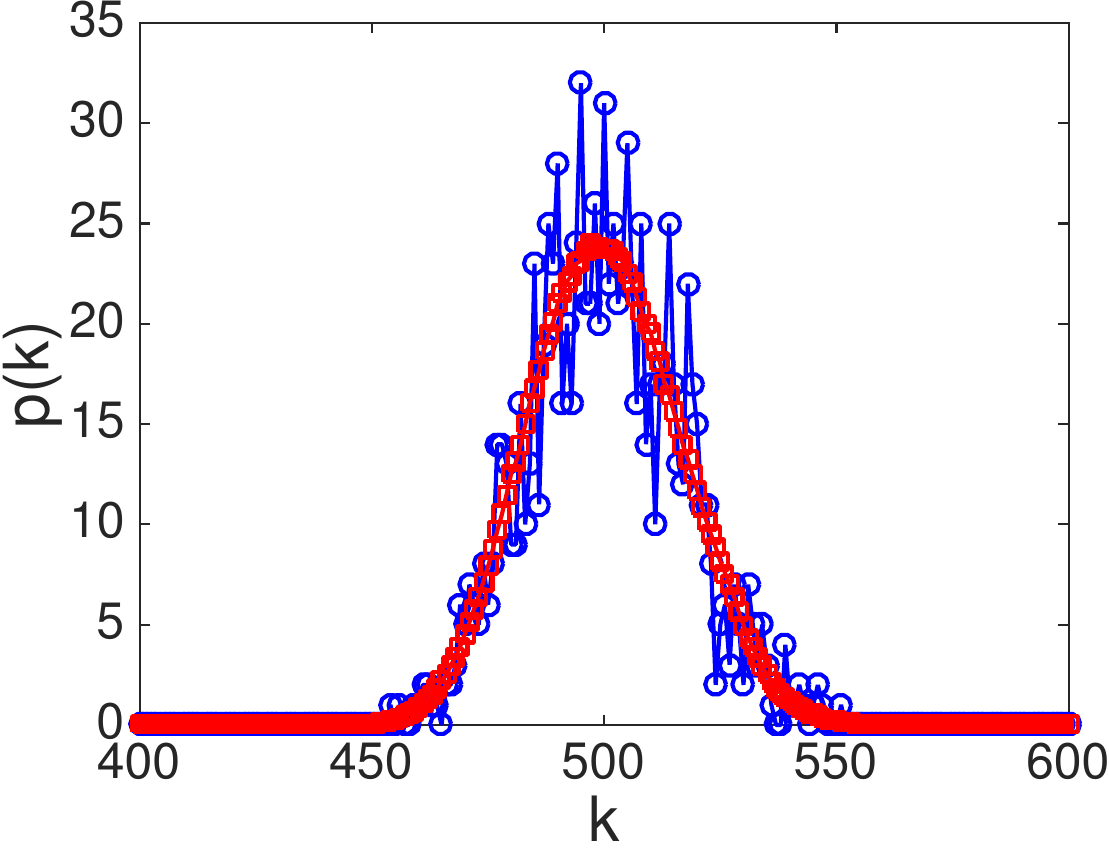} 
\end{tabular}
\end{centering}
\caption{Random walkers are made to evolve on a Erdos-Renyi network  made by $\Omega=1000$. The probability $p$ for a link to exist between any two pairs of nodes is set to $0.5$. The exact (blue circles) and reconstructed (red squares) profiles are displayed. From left to right: $s=510$ [panel (a)], $s=530$ [panel (b)]$, s=550$ [panel (c)]$, s=600$ [panel (d)] .}
\label{figSI:Fig0}
\end{figure*}

\begin{figure*}[ht!]
\begin{centering}
\includegraphics[scale=0.4]{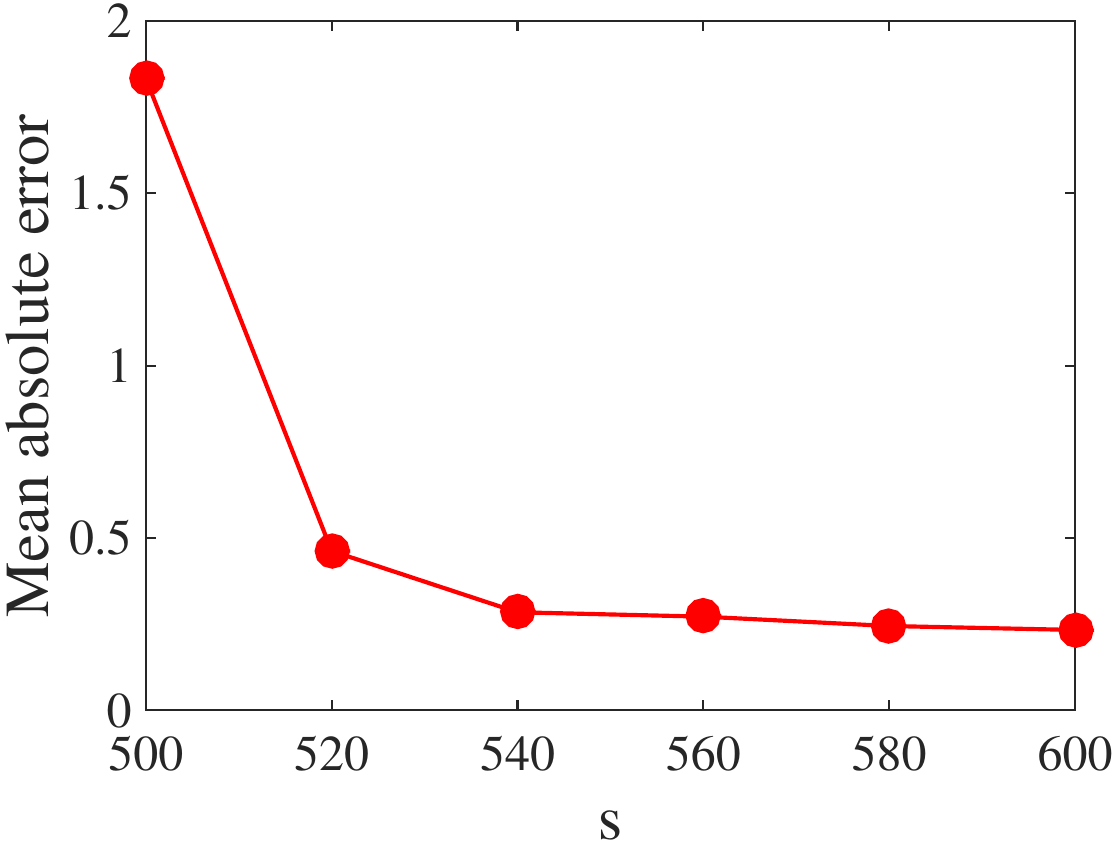} 
\end{centering}
\caption{Error in the reconstruction vs. the parameter $s$. The error is defined as the sum over $k$ of the difference, in absolute value, between the reconstructed and true $p(k)$ profiles divided by 
the number of nodes $\Omega$. For the parameters refer to Figure SI \ref{figSI:Fig0}.}
\label{figSI:Fig01}
\end{figure*}

In Figure SI \ref{figSI:Fig0} the convergence of the algorithm of reconstruction is assessed, when varying the parameter $s$. The reconstructed profile converges steadily towards the correct distribution. In all cases, the number of nodes of the network is correctly estimated. In the example displayed an Erdos-Renyi network is assumed with $\Omega=1000$ and $p=0.5$, so as to make contact with the results reported in the main body of the paper. The monotonous convergence of the algorithm can be also appreciated in Figure SI \ref{figSI:Fig01}, where the sum (divided by $\Omega$) over $k$ of the difference (in absolute value) between exact and approximated distributions is plotted against $s$.

\section{Stochastic simulations vs. deterministic equilibrium.}

As stated in the main body of the paper, when a finite population of walkers is made to evolve on a network the dynamics is intrinsically stochastic. In this section, we will discuss the validity of the mean-field 
equilibrium (a key ingredient of the devised inverse problem), when the algorithm for the reconstruction of the underlying network is feeded on stochastic trajectories. To this end we produce independent realizations of the stochastic dynamics (as ruled by  the master equation (1) in the main body of the paper) via the celebrated Gillespie algorithm. We will in particular implement an asynchronous update of the system dynamics, to prevent instantaneous collision of individual agents simultaneously heading towards an isolated vacancy.  

\begin{figure*}[ht!]
\begin{centering}
\begin{tabular}{lll}
\textbf{a} & \textbf{b} & \textbf{c} \\
\includegraphics[scale=0.2]{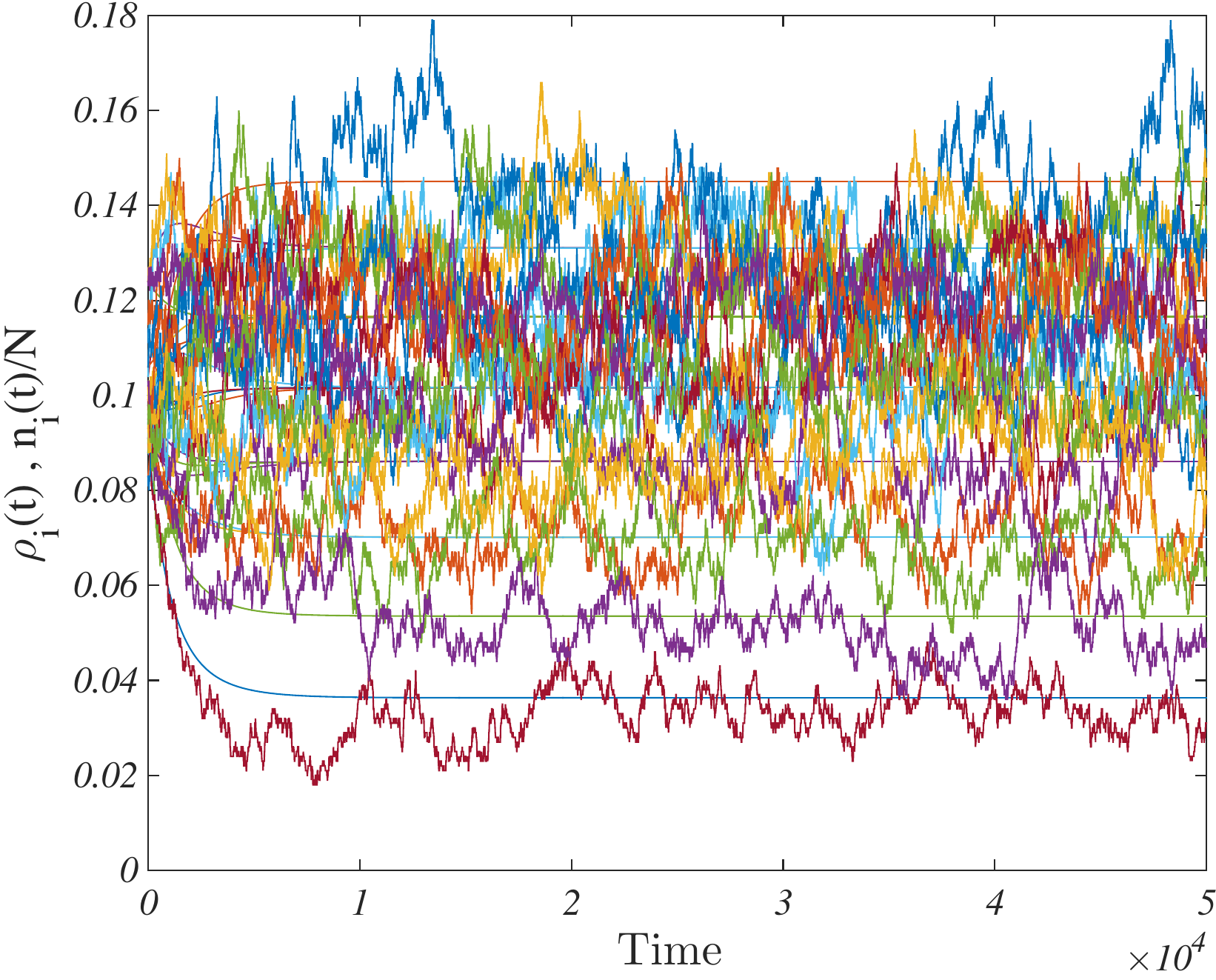} & \includegraphics[scale=0.2]{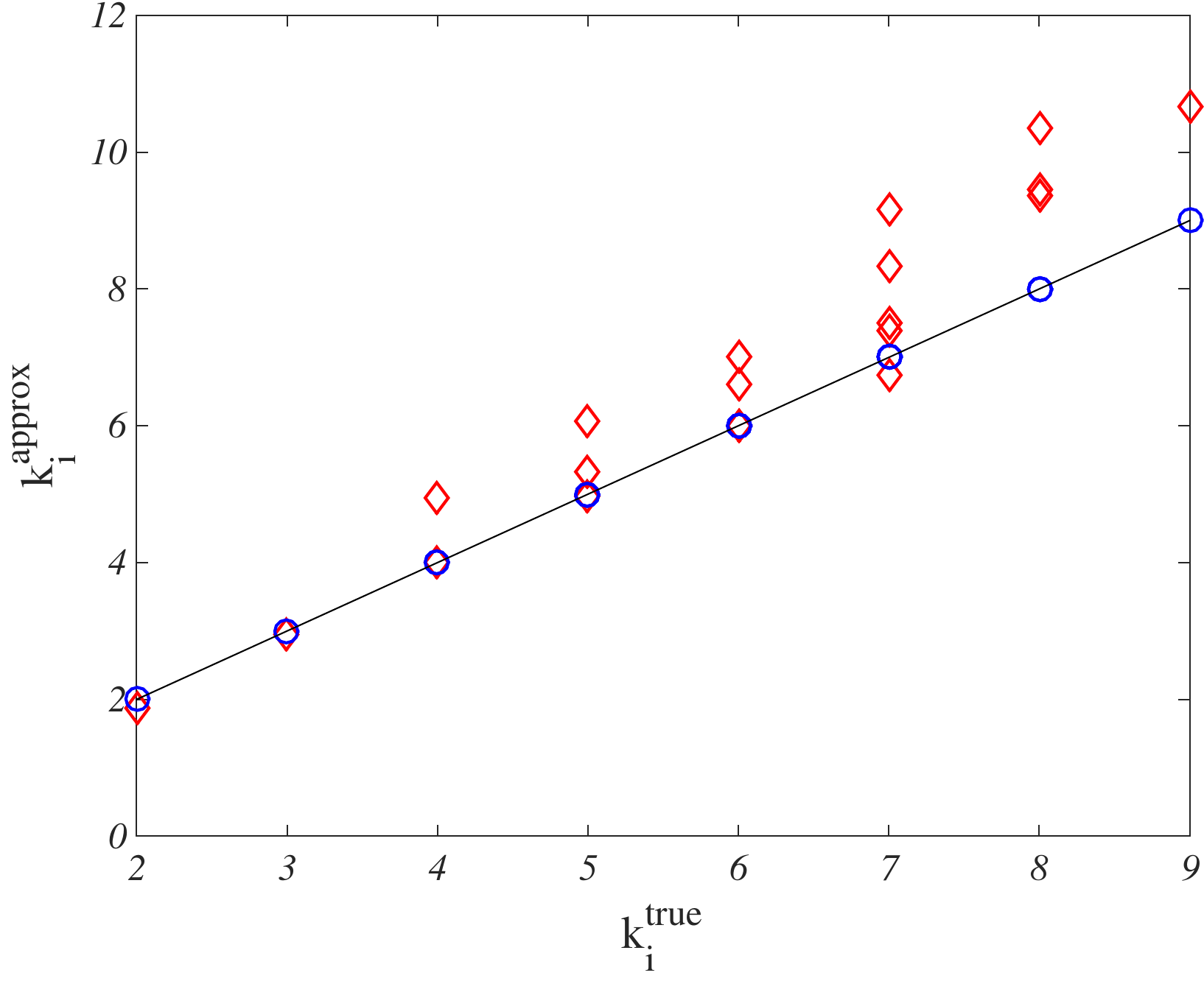} & \includegraphics[scale=0.2]{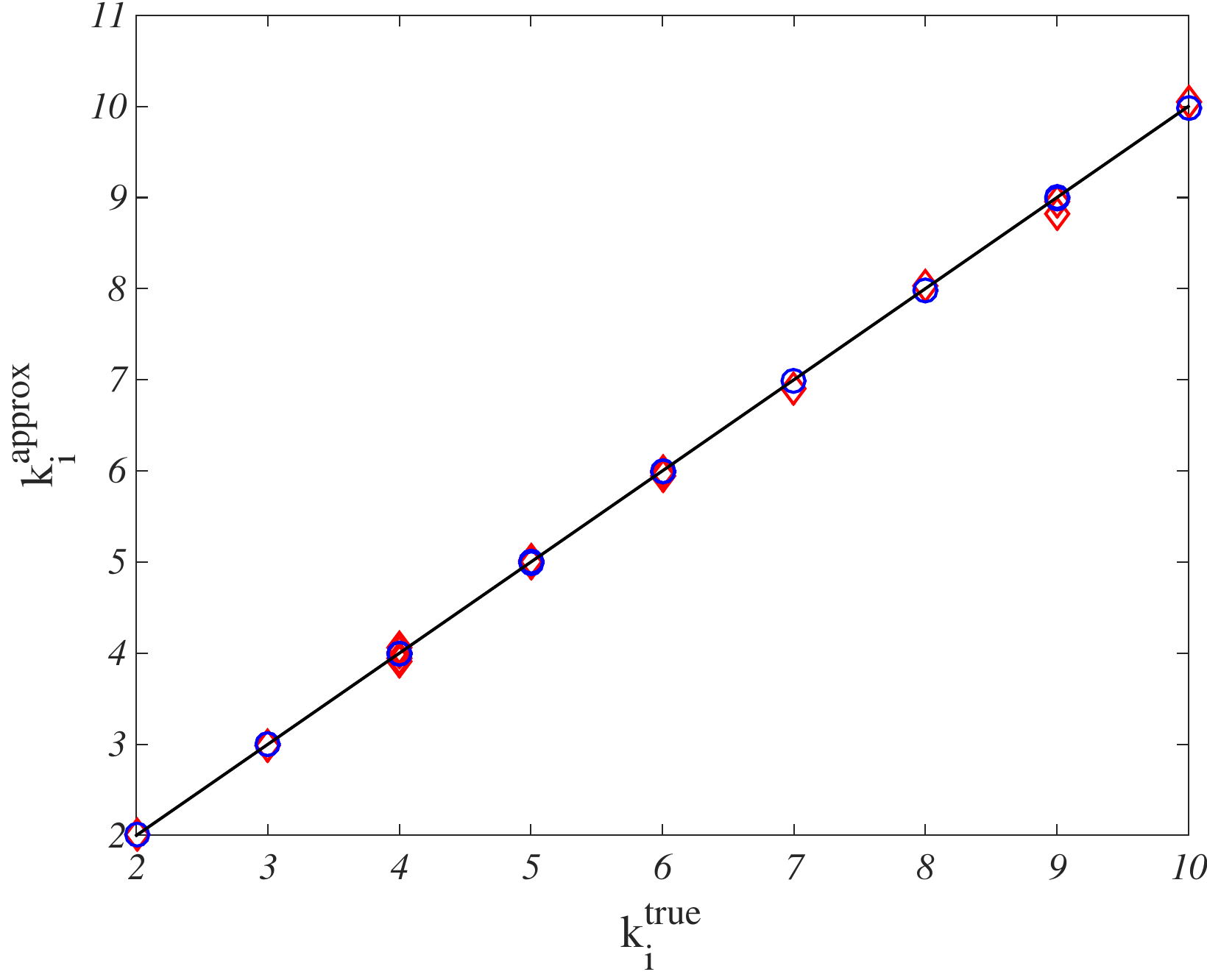} 
\end{tabular}
\end{centering}
\caption{Degree reconstruction in the stochastic framework. Panel (a): time evolution of $n_i(t)/N$ for a single run of the Gillespie algorithm.   The smooth solid curves stand for the corresponding deterministic solutions.
Panel (b):  $k_i^{approx}$ vs. $k_i^{true}$.  Red diamonds refer to the estimates obtained via stochastic simulations, as displayed in panel (a). Blue circles are instead obtained by operating in the deterministic limit. 
Panel (c): $k_i^{approx}$ vs. $k_i^{true}$. The parameter $a$ is calculated as the average of different estimates, each obtained by sampling  individual stochastic trajectories on a given node. Here, we considered $100$ trajectories of the same time duration as those displayed in  panel (a). A Erdos-Reny random network made by $\Omega=20$ is assumed, with probability for a link to be established equal to $p=0.3$. The initial conditions for the stochastic simulations are drawn from a binomial distribution with $N=1000$.  Here, $\beta=0.1 \Omega$.}
\label{figSI:Fig1}
\end{figure*}

To perform the analysis we consider a Erdos-Renyi random network made by $\Omega=20$ nodes. The probability to draw a link between two randomly selected nodes is set to $p=0.3$. Initial conditions for the population of discrete agents are drawn form  a binomial distribution with $N=1000$, which hence plays the role of the carrying capacity. Further we set $\beta=0.1 \Omega$. In Figure SI \ref{figSI:Fig1}(a), $n_i/N$ is plotted versus time and compared to the solution of the mean field equation, $\rho_i(t)$, subject to initial conditions $\rho_i(0)=n_i/N$. Averaging over different realizations of the stochastic dynamics yields a progressive convergence towards the deterministic solution. To perform the network reconstruction, one needs to compute the parameter $a$ which enters equation (4), in the main body of the paper. It is crucial that $a$ is correctly determined for the method to converge to the sought distribution $p(k)$. To test how the stochastic fluctuations impact on the procedure, we refer to the simulations displayed in  Figure SI \ref{figSI:Fig1}(a) and select at random one node $j$, whose connectivity $k_j$, is supposed to be known. We then estimate the average node occupancy $\langle n_j \rangle$, normalized to the carrying capacity $N$, to yield the number density $\langle n_j \rangle/N$. This latter is used to obtain an approximated estimate for the parameter $a$: the degree $k_i$ of all nodes $i$ can be hence retrieved and compared to the corresponding true values (which can be straightforwardly computed in this artificial example). This check provides an indirect test for the accuracy of the reconstruction, as accessing a reliable and robust estimate for the parameter $a$ is mandatory for the method to converge smoothly.  As displayed in Figure SI \ref{figSI:Fig1}(b), predictions based on the stochastic input show important deviations from the correct entries, which are instead adequately captured within the idealized deterministic framework. To overcome this limitation,  it is however enough to improve on the estimate of $a$ by averaging over many independent realizations of the stochastic dynamics (or equivalently by increasing the time of acquisition of the signal), as sampled on the individual node $j$. In Figure SI \ref{figSI:Fig1}(c), the connectivity $k_i$ obtained by using an improved estimate of $a$ (which combines $100$ independent stochastic runs) is plotted and show excellent agreement with the corresponding values found under the deterministic scenario.   

\section{Recovering the information from one node (for longer time) or from several nodes (for shorter time)?}

The aim of this section is to critically compare in terms of performances three alternative approaches to the investigated problem. We could in fact imagine to retrieve the information needed for the reconstruction protocol from just one node (as it is done all over the paper) and improve on the statistics by combining data obtained from $L$ different windows of acquisition, with identical time duration $w_{size}$. Alternatively, we could simultaneously gather information from $L$ nodes sampling just one time segment of fixed extension $w_{size}$. In both cases the available information is of the same order and relates to the quantity $L\times w_{size}$. A third possibility would be to sample the dynamics from just one isolated  node, for a single time window of length $L\times w_{size}$. Note that the window size is expressed in time units, hence a given window can contain a different number of events (jumps or Gillespie steps), as Gillespie times are not uniform.

More precisely, the different  sample strategies as identified above can be explained as follows:
\begin{itemize}
\item[(1)] solve the model up to a time $t_{fin}$ and, after a sufficiently long transient, measure $n_i(t)$ (or $\rho_i(t)$) for $i\in\{1,\dots,L\}$ nodes.  Then, compute the time average for the relevant state variables over the sampled window, namely $\langle n_i(t)\rangle_{w}$ (when operating under the deterministic scenario, this latter step is not necessary, assuming $\rho_i(t)$ has reached the asymptotic state). From the collection of $L$ values $\langle n_i(t)\rangle_{w}$ compute the corresponding estimate of $a_i$ as:
\begin{equation*}
a_i=\frac{\langle n_i(t)\rangle_{w}/N}{k_i (1-\langle n_i(t)\rangle_{w}/N)}\, ,
\end{equation*}
where $k_i$ is the degree of the $i$--th node and $N$ the associated carrying capacity. Finally,  define the average $\langle a^{(1)}\rangle=\sum_i a_i/L$ and  use it to reconstruct the network following the usual scheme.
\item[(2)] solve the model up to a time $t_{fin}$ and, after a sufficiently long transient, measure $n_1(t)$ (or $\rho_1(t)$). Here, the selection of node $1$ is arbitrary. Then compute the time average over the $L$ windows, $\langle n_1(t)\rangle_{w_j}$ (again, this step is not necessary in the deterministic framework, if $\rho_1(t)$ already reached its asymptotic state). From the $L$ estimates $\langle n_1(t)\rangle_{w_j}$ compute $a_i$ as:
\begin{equation*}
a_i=\frac{\langle n_1(t)\rangle_{w_j}/N}{k_i (1-\langle n_1(t)\rangle_{w_j}/N)}\, .
\end{equation*}
Finally define the average $\langle a^{(2)}\rangle=\sum_i a_i/L$ and use it to reconstruct the network, based on the protocol illustrated and tested in the main body of the paper..
\item[(3)] solve the model up to a time $t_{fin}$ and, after a sufficiently long transient, measure $n_1(t)$ (or $\rho_1(t)$). Again, the node $1$ is arbitrarily selected. Then compute the time average over the large window, $\langle n_1(t)\rangle_{W}$ (in the deterministic case this step can be omitted, provided $\rho_1(t)$ has relaxed to its asymptotic equilibrium). Then compute $a^{(3)}$ by means of:
\begin{equation*}
a^{(3)}=\frac{\langle n_1(t)\rangle_{W}/N}{k_i (1-\langle n_1(t)\rangle_{W}/N)}\, .
\end{equation*}
\end{itemize}

\begin{figure*}[ht!]
\begin{centering}
\begin{tabular}{ll}
\textbf{a} & \textbf{b} \\
\includegraphics[scale=0.3]{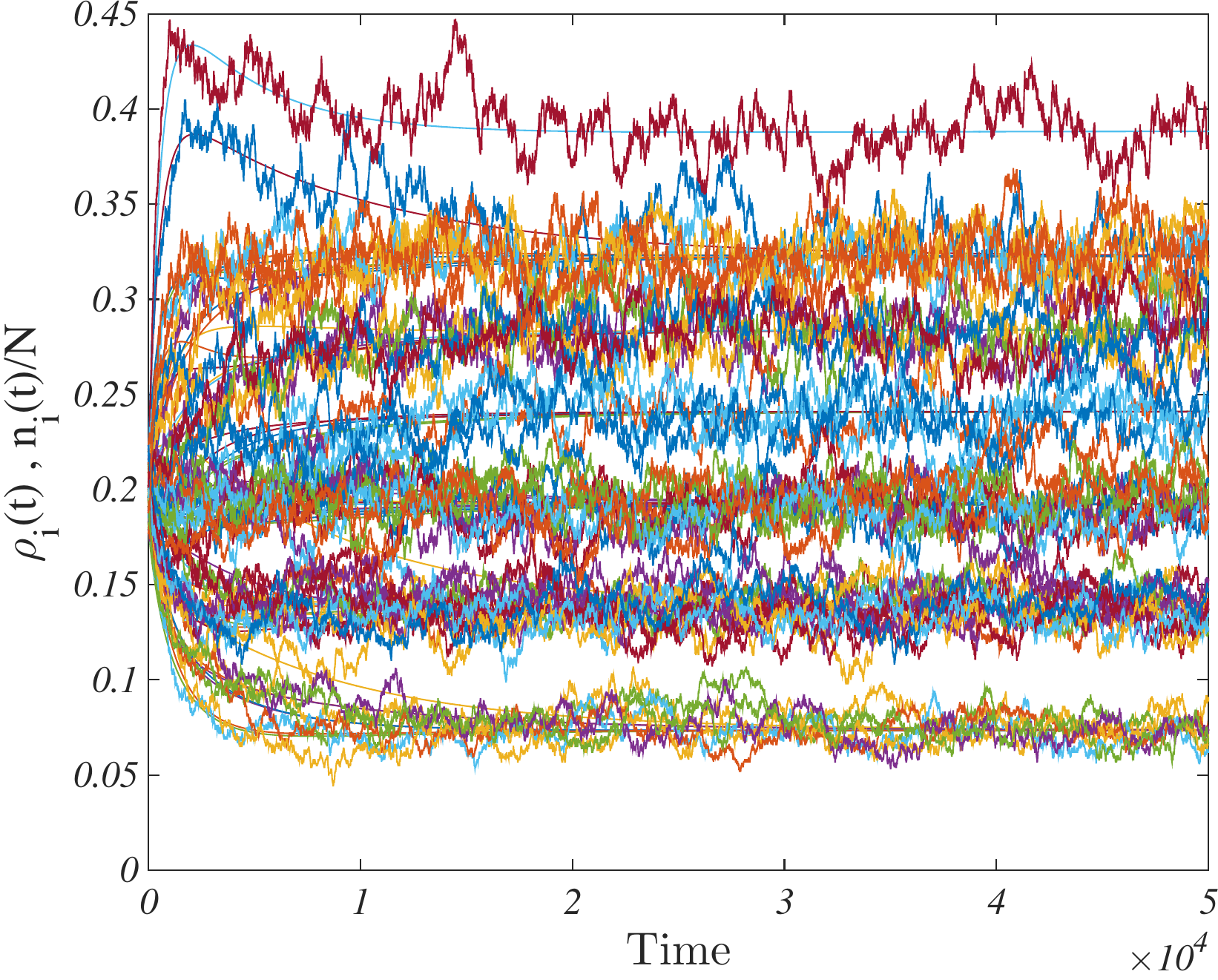} & \includegraphics[scale=0.3]{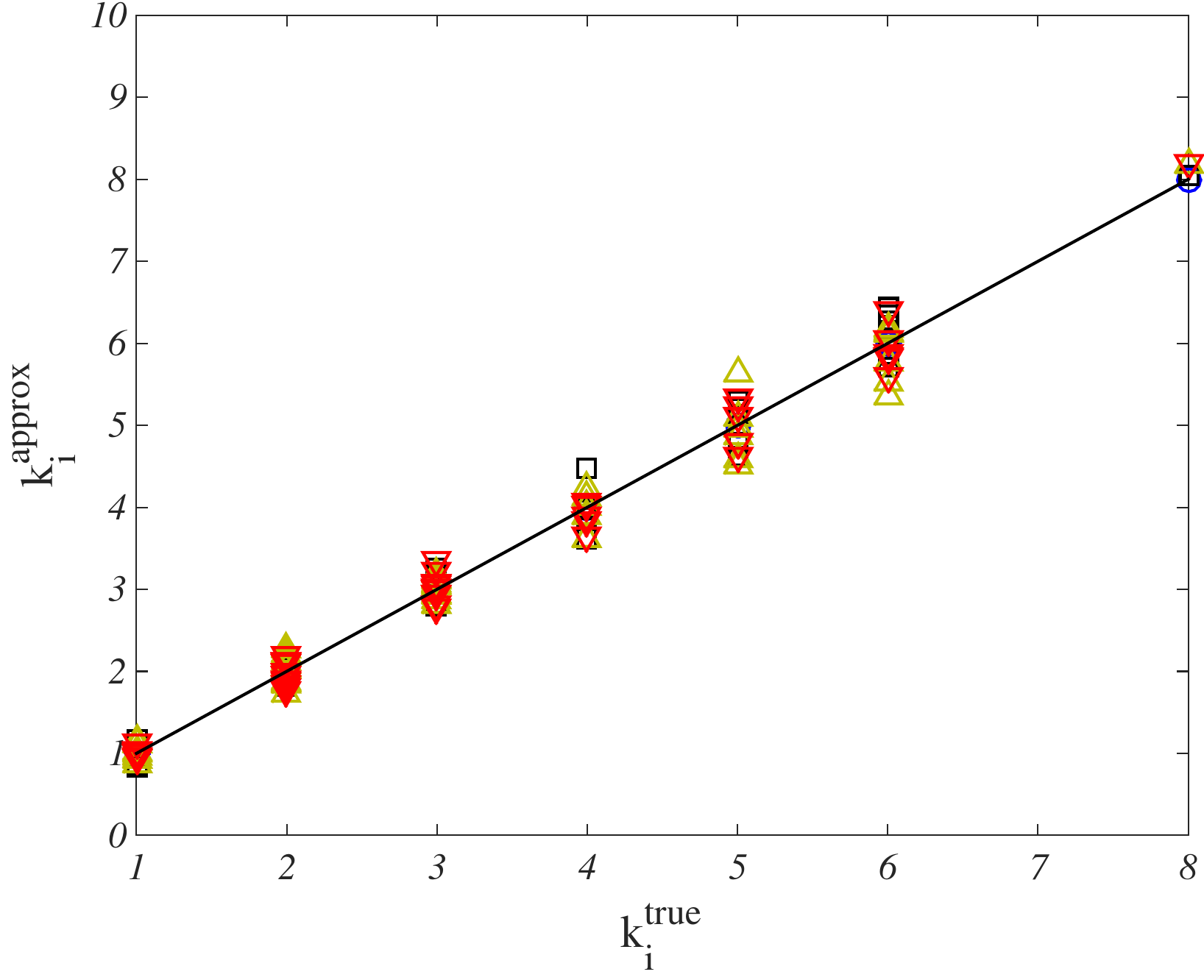} 
\end{tabular}
\end{centering}
\caption{Panel (a): stochastic vs. deterministic trajectories. Panel (b):  $k_i^{approx}$ vs. $k_i^{true}$ as obtained by using the different strategies  outlined in the text.  Black squares refer to strategy (1), green upper triangles to strategy (2) and red down triangles to strategy (3). Blue circles (barely visible) are obtained under the deterministic viewpoint. Simulations are performed for a Erdos-Renyi random network made by $\Omega=50$ nodes and $p=0.06$. Initial conditions are drawn from a binomial distribution with $N=1000$. Further, $\beta=0.2 \Omega$. Here, $L=10$ and $w_{size}=500$.  }
\label{figSI:Fig2}
\end{figure*}

Working under the deterministic scenario, the three approaches are equivalent: there is no gain from acquiring several measures from distinct nodes or, alternatively, using multiple time windows (provided the dynamics is close enough to the asymptotic state), since the reconstructed $a$ is independent from the selected node and solely affected by the total amount of mass (or total number of agents) flowing on the network. In Figure SI \ref{figSI:Fig2}, we test
the three different approaches when fluctuations are allowed for, namely when operating under a stochastic perspective. Simulations are carried out for a Erdos-Renyi random network made by $\Omega=50$ nodes and $p=0.06$.
Initial conditions are drawn from a binomial distribution with $N=1000$. Further, $\beta=0.2 \Omega$. Here, $L=10$ and $w_{size}=500$. In panel (a) we show the time evolution of respectively the stochastic (wiggling curves) and the deterministic (smooth solid lines) systems:  $n_i(t)/N$ is plotted against the stochastic time (not rescaled by $N$). In panel (b) we compare the degrees reconstructed via the different strategies as illustrated above: black squares refer to strategy (1), green upper triangles to strategy (2) and red down triangles to strategy (3). Blue circles (barely visible) stands for the values obtained in the mean field limit. In this latter case, the three strategies amount to the same, provided the asymptotic equilibrium has been eventually attained.   

In conclusion, different approaches returns equally accurate estimates.  Method (1) only requires accessing the stochastic trajectories over a relatively short time window (in addition to a transient, which needs however to be accounted for in all considered methods), while approaches (2) and (3) necessitate that the system is eventually followed over a longer time period.  In conclusion, if the integration scheme is costly,  strategy (1) is preferable.

\section{On the convergence of the moments of the distribution $p(k)$: a rule of the thumb to determine the largest connectivity value.}

\begin{figure*}[ht!]
\begin{centering}
\begin{tabular}{ll}
\textbf{a} & \textbf{b} \\
\includegraphics[width=0.4\textwidth]{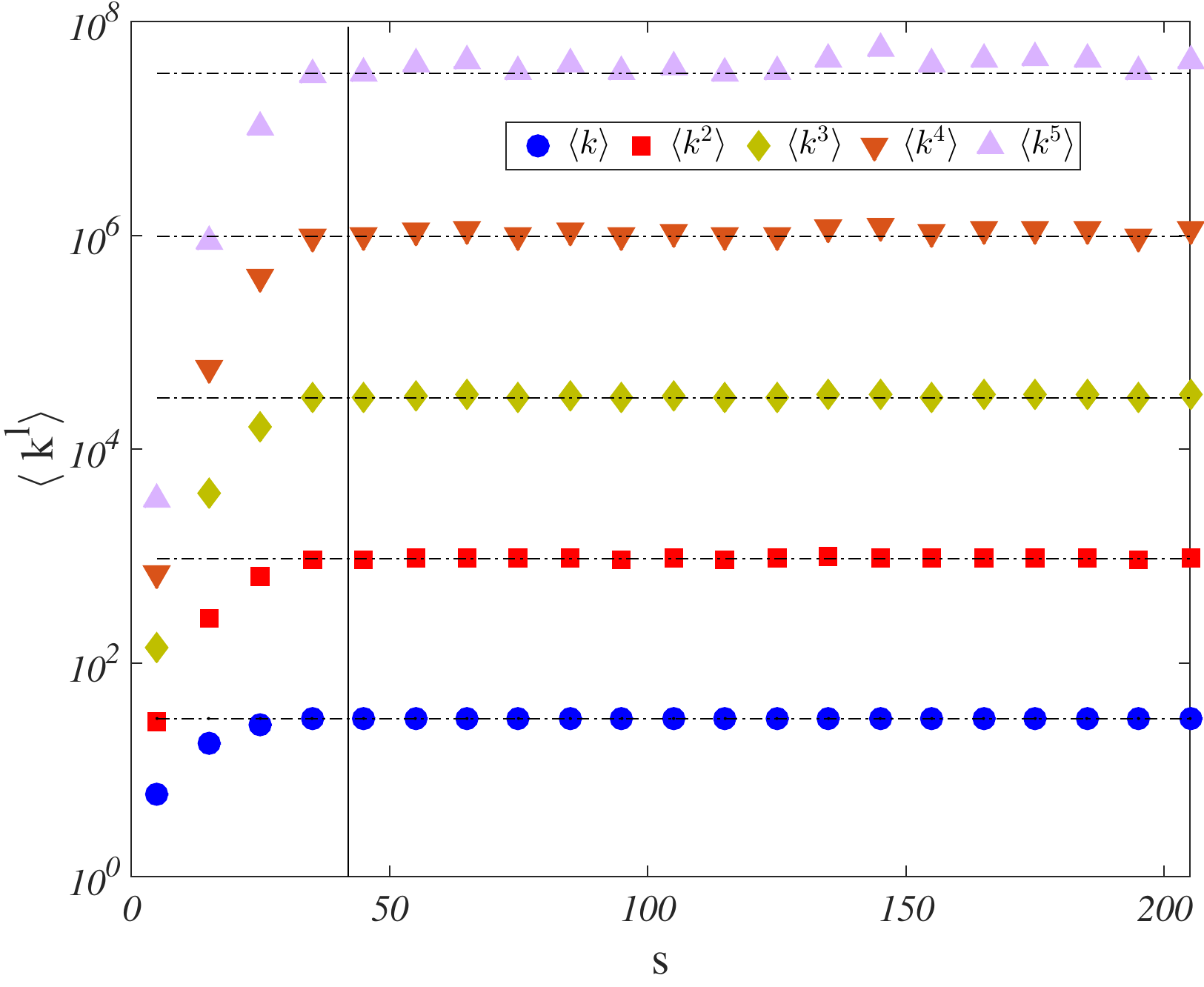} &  \includegraphics[width=0.4\textwidth]{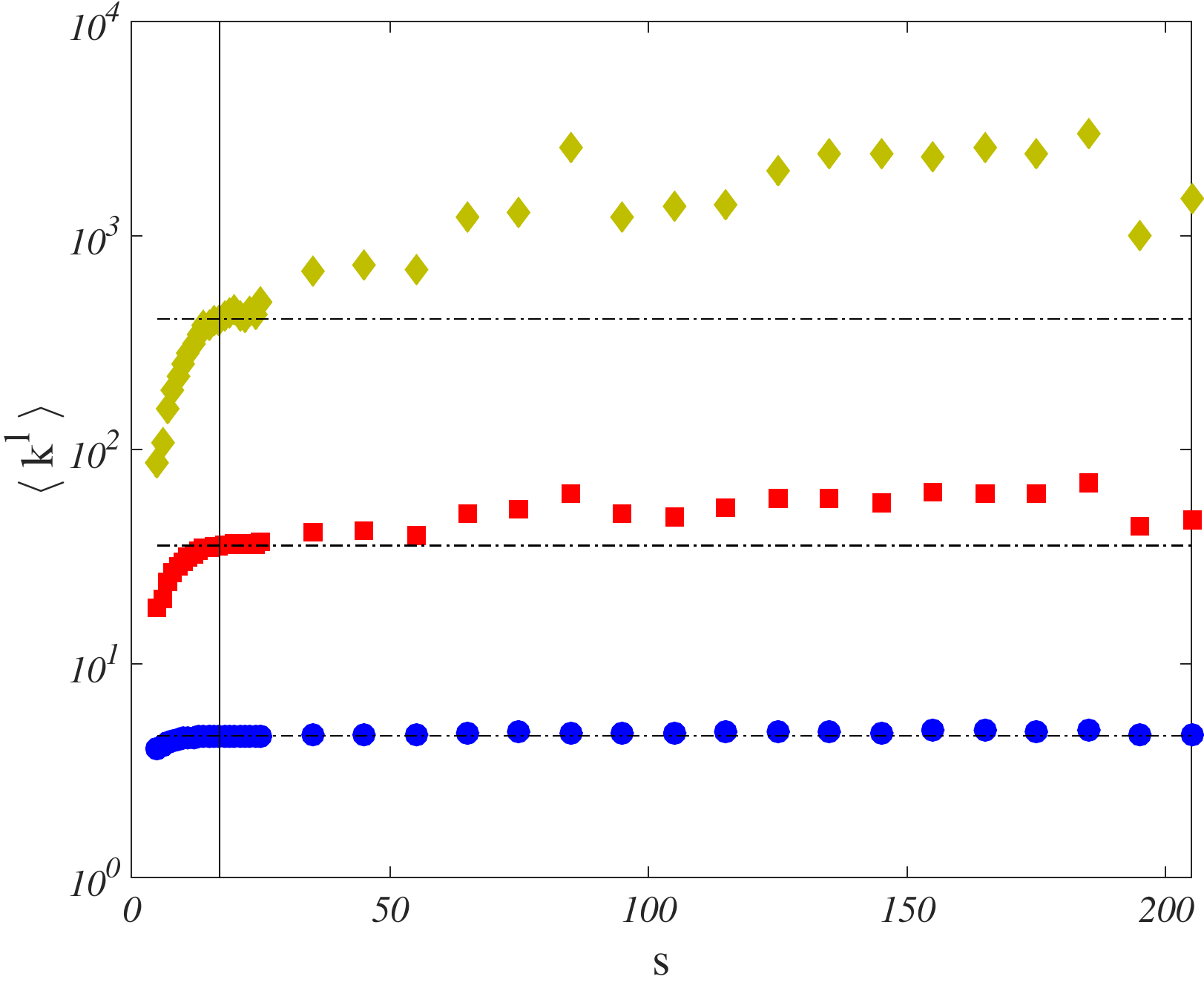}
\end{tabular}
\end{centering}
\caption{Panel (a): $\langle k^l \rangle$ is plotted against $s$, for $l=1, \dots, 5$ (from bottom to top). The moments are calculated from the reconstructed $p(k)$, working at different values of $s$. The horizontal dot-dashed lines identify the exact values for the considered moments. The vertical solid line is traced at $s=k_{max}$ and correctly identifies the saturation point for all computed moments. Here, a Erdos-Renyi network made of $\Omega=100$ nodes and $p=0.3$ is considered. Panel (b): $\langle k^l \rangle$ is plotted against $s$, for $l=1, \cdots, 3$ (from bottom to top). The moments are calculated from the reconstructed $p(k)$, working at different values of $s$. The horizontal dot-dashed lines stand for the exact values of the computed moments. The vertical solid line is traced at $s=k_{max}=17$ and identifies the turning point, as displayed by the empirical moments plotted against $s$.
Here, the Karate Club network is considered. }
\label{figSI:Fig3}
\end{figure*}

Let us start by considering an Erdos-Reny network made of $\Omega=100$ nodes and $p=0.3$. In this case the maximum degree of the network is found to be $k_{max}=42$. We perform the network reconstruction assuming different values of $s$ and use the estimated $p(k)$ to compute the associated moments $\langle k^l \rangle = \sum_k k^l p(k)$ for $l=1, \dots, 5$. Results are displayed in Figure SI \ref{figSI:Fig3}, panel (a), where $\langle k^l \rangle$ si plotted against $s$. As it can be readily appreciated by visual inspection, when increasing $s$ the reconstructed moments quickly converge towards the exact values, here identified by the horizontal dot-dashed lines. The convergence is attained when $s$ is of the same order of $k_{max}$ (see vertical solid line), as we have anticipated in the main body of the paper. Monitoring the evolution of the first few moments as function of the sampling parameter $s$, could hence provide a reliable strategy to determine the a priori the unknown quantity $k_{max}$. On the other hand, it enables one to quantify the smallest number of replica $s$ that need to be analyzed for a successful computation of the sought distribution, so returning a self-consistent stopping criterion.

In Figure SI \ref{figSI:Fig3}, panel (b), we report a similar analysis for the Karate Club network, yielding identical conclusions.

\end{document}